
\magnification=1200
\baselineskip=6mm
{\nopagenumbers
\noindent
\null\vskip 3truecm
\centerline
{\bf THE GEODESIC MOTION ON GENERALIZED TAUB-NUT }
\centerline{\bf GRAVITATIONAL INSTANTONS}

\vskip 2truecm
\centerline{Mihai Visinescu~\footnote{*}{E-mail address: ~~MVISIN@ROIFA}}
\centerline{Department of Theoretical Physics, }
\centerline{Institute of Atomic Physics, P.O.Box MG-6}
\centerline{Magurele,Bucharest,Romania}

\vskip 3truecm
\centerline{\bf Abstract}

A class of generalized Taub-NUT gravitational instantons is
reported in five - dimensional Einstein gravity coupled to a
non-linear sigma model. The geodesic dynamics of a spinless
particle of unit mass on these static gravitational instantons
is studied. This is accomplished by finding a generalized
Runge-Lenz vector. Unlike the Kepler problem, or, the dynamics
of a spinless particle on the familiar Taub-NUT gravitational
instantons, the orbits are not conic sections.

\vskip 1truecm

\vfil\eject
}
\pageno=1
{\bf 1. INTRODUCTION}
\vskip 1truecm

The classical equations of sourceless five-dimensional general relativity
admit various stationary solutions among which the well-known Kaluza-Klein
multi-monopoles [1]. The Kaluza-Klein monopole was obtained by embedding
the Taub-NUT gravitational instanton into five-dimensional theory,
adding the time coordinate in a trivial way. Its line element is
expressed as
$$ds^2_5=-dt^2+ds^2_4$$
$$=-dt^2+V^{-1}(r)~[dr^2+r^2d\theta^2+
r^2\sin^2\theta~d\varphi^2~]+V(r)~[dx^5+{\bf A}({\bf r}~)\cdot
d{\bf r}~]~^2  \eqno(1.1)$$
where ${\bf r}$  denotes a three-vector ${\bf r} = (r,\theta,\varphi)$
and the gauge field ${\bf A}$ is that of a monopole
$$A_r=A_\theta=0,~~~A_\varphi=4m(1-\cos\theta)$$
$${\bf B}= rot ~{\bf A}~={4m{\bf r}\over r^3}~~. \eqno(1.2)$$

The function $V(r)$ is
$$V(r)=~\left(1~+~{4m\over r}\right)^{-1} \eqno(1.3)$$
and the so called NUT singularity is absent if $x^5$ is periodic
with period $16\pi m$ [2].

Remarkably the same object has re-emerged in the study of
monopole scattering. In the long distance limit, neglecting
radiation, the relative motion of two monopoles is described by
the geodesics  of the space (1.1) [3,4]. Slow Bogomolny - Prasad -
Sommerfield monopoles move along geodesics in a
four-dimensional curved space with the line element $ds^2_4$.
The dynamics of well-separated monopoles is completely soluble,
but not trivial [3-8].The problem of geodesic motion in
this metric has therefore its own interest, independently of
monopole scattering.

In this paper we use similar methods to investigate the geodesic
motion in the space of a Kaluza - Klein monopole in the presence
of a scalar sigma field coupled to the metric tensor
field. It was noted several years ago by Omero and Percacci [9]
and Gell-Mann and Zwiebach [10] that the nonlinear sigma-model
can be used to induce the space-time compactification in
Kaluza-Klein theory. This mechanism presents many interesting
features among others being the absence of a cosmological term
at the classical level.

The model we shall discuss consists of Einstein gravity in 5
dimensions coupled to a nonlinear sigma-model
$$S~=~-{1\over 16\pi G_k}\int_{M^5} d^5 x\sqrt{-g_5}\left[~R_5
-{2\over\lambda^2}g^{AB}~{\partial \Phi\over\partial x^A}~
{\partial \Phi\over\partial x^B}\right] \eqno(1.4)$$
where $R_5$ is the five-dimensional scalar of the metric
$g_{AB}$ of the manifold $M^5$ with the signature $-++++$ . Upper case Latin
letters $A,B,C...$ denote five-dimensional indices $0,1,2,3,5$ and
$\lambda^2$ is a constant giving the strength of the
self-coupling of the scalar fields.

The corresponding Euler-Lagrange equations are the Einstein equations
$$R_{AB}~=~{2\over \lambda^2}~{\partial \Phi\over\partial x^A}~
{\partial \Phi\over\partial x^B} \eqno(1.5)$$
and an aditional one which prescribes that $\Phi$ is a harmonic
map [9,11]. In [11], using the Einstein equations, we came to
the conclusion that $\Phi$ is harmonic if the map is a
differentiable submersion almost everywhere (see also [12]).
Therefore a very general class of solutions of the model is
given by submersions $\Phi~:~M^5~\rightarrow ~B~$  satisfying
Einstein equations (1.5) where  $B$  is the manifold in which the
scalar field  $\Phi$  takes values [11,13,14].

Recently we shown [15,16] that in the model there exist static
solutions of multi-monopole type. Inspired by the previous
results [1,2,17] we used the static ansatz (1.1) for the
five-dimensional metric. For a static configuration of $N$
monopoles located at ${\bf a}_l$ we found that $V$ is a
function of ${\bf r}$ through the variable
$$\rho~=~1~+~\sum_{l=1}^N {4m_l\over \vert {\bf r} -{\bf a}_l
\vert } \eqno(1.6)$$
satisfying the following equation
$${1\over V}~{d^2\over d\rho^2}~\left({1\over V}\right)~-\left[
{d\over d\rho}~\left({1\over V} \right)\right]^2~+~1~=~0~~.\eqno(1.7)$$

This equation can be solved explicitely and the general
solutions are
$$V~=~{\alpha\over \sin(\alpha\rho~+~\beta)} \eqno(1.8)$$
$$V~=~{\alpha\over \sinh(\alpha\rho~+~\beta)} \eqno(1.9)$$
$$V~=~{1\over \pm\rho~+~\beta} \eqno(1.10)$$
where $\alpha ,~\beta$ are integration constants and the scalar field $\Phi$
can be determined from the differential equation :
$${d^2\over d\rho^2}~\left(~{1\over V}~\right)~=~-{4\over
\lambda^2}~{1\over V}~\left({d\over d\rho }~\Phi~\right)^2 \eqno(1.11).$$

In the present paper we shall restrict ourselves to the monopole
solution $(N~=~1)$ located at the origin of the system of
coordinates $({\bf a}_1~=~{\bf 0}~)$. In the next section we
shall investigate the geodesic motion in the proper space of the
Kaluza-Klein monopole in the presence of a scalar field. Of
course , for a vanishing scalar field (i.e. in the limit
$\alpha~\rightarrow ~0 , \beta~\rightarrow ~0 ,
\lambda~\rightarrow ~0 $ in eqs. (1.8)-(1.11) we recover the
standard monopole solution (1.2), (1.3) with the dynamics
described in papers [3-8].

In spite of the presence of the scalar field $\Phi$ , there are
many similarities with the dynamics of the standard monopoles in
an euclidean Taub-NUT metric. However there are some notable
differences and the most important one is a generalization of
the unexpected Runge-Lenz vector which appears in the monopole
dynamics. Unlike the Kepler problem, or, the dynamics of a
spinless particle on the familiar Taub - NUT gravitational
instantons, the orbits are not conic sections.

The paper ends with some concluding comments and with a
perspective on various conceivable generalizations.
\vskip 2truecm
{\bf 2.CLASSICAL DYNAMICS}
\vskip 1truecm

As stated above we shall consider that the space-time
compactification is induced by a scalar field in the form of a
nonlinear sigma- model coupled to gravity as in eq. (1.4). We
shall make the ansatz (1.1) for the metric with the function $V$
having one of the forms (1.8)-(1.10) with
$$\rho~=~1~+~{4m\over r} \eqno(2.1)$$
corresponding to the monopole configuration (1.2) . It is
convenient to make the coordinate transformation
$$4m~(\psi~+~\varphi )~=~-x^5 \eqno(2.2)$$
with $0~\leq ~\psi~< ~4\pi $ , which converts the four-dimensional line
element $ds_4$ into
$$ ds^2_4~=~V^{-1}(r)~[dr^2~+~r^2d\theta ^2~+~r^2\sin^2\theta~d\varphi^2~]~
+~16m^2V(r)~[d\psi~+~\cos\theta d\varphi~]^2$$
$$=g_{\mu\nu}~dx^\mu dx^\nu~~~~~~,~~~~~(\mu ,\nu~=~1,2,3,5)
\eqno(2.3) $$

Spaces with a metric of the form given above have an isometry
group $SU(2)\times U(1) $. The four Killing vectors are
$$\xi_1~=~\sin\varphi~{\partial\over \partial\theta}~+~\cos\varphi~\cot\theta~
{\partial\over \partial\varphi}~-~{\cos\varphi\over
\sin\theta}~{\partial\over \partial\psi}$$
$$\xi_2~=~-\cos\varphi~{\partial\over
\partial\theta}~+~\sin\varphi~\cot\theta~{\partial\over \partial\varphi}~
-~{\sin\varphi\over \sin\theta}~{\partial\over \partial\psi}$$
$$\xi_3~=~{\partial\over \partial\varphi}$$
$$\xi_5~=~{\partial\over \partial\psi}~~.  \eqno(2.4)$$

$\xi_5$ , which generates the $U(1)$ of $\psi$ translations ,
commutes with the other Killing vectors. In turn the remaining
three vectors obey an $SU(2)$ algebra with
$$[~\xi_1~,~\xi_2~]~=~-\xi_3~~~,~etc...  \eqno(2.5)$$

This can be contrasted with the Schwarzschild space-time where
the isometry group at spacelike infinity is $SO(3)~\times ~U(1)$.
 This illustrates the essential topological character of the
magnetic monopole mass [18]. On the other hand, it stand to
reason that the static metric (1.1) has $\xi_0~=~{\partial\over
\partial t} $ as an additional Killing vector which
generates the $U(1)$ of time translations and commute with the
other four Killing vectors (2.4).

The geodesic motion of a spinless particle of unit mass in (2.3)
is described by the Lagrangian
$$L~=~{1\over2}~g^{\mu\nu}~\dot{x}_\mu~\dot{x}_\nu$$
$$=~{1\over2}~\left[ {1\over V(r)}~{\dot{\bf r}}^2~+~(4m)^2~V(r)(\dot{\psi}~+
{}~\cos\theta~\dot{\varphi})^2~\right] \eqno(2.6)$$
where the dot refers to differentiation with respect to proper
time.

To the two cyclic variables $\psi$ and $t$ are associated the
conserved quantities
$$q~=~(4m)^2~V(r)~(~\dot{\psi}~+~\cos \theta~\dot{\varphi}~)\eqno(2.7)$$
$$E~=~{1\over2V(r)}~\left[~{\dot{\bf
r}}^2~+~\left({q\over4m}\right)^2\right]~=~
{1\over2}~\left[~V(r)~{\bf p}^2~+~{1\over
V(r)}~\left({q\over4m}\right)^2\right] \eqno(2.8)$$
where
$${\bf p}~=~{1\over V(r)}~{\dot{\bf r}}\eqno(2.9)$$
is the ''mechanical momentum'' which is only part of the
momentum canonically conjugate to ${\bf r}$ . For the monopole
scattering $q$ and $E$ are interpreted as ''relative electric
charge'' and energy, respectively.

The equation of motion for ${\bf p}$ is
$${\bf p}~=~{1\over2}~{\vec{\bf \nabla}}~\left({1\over
V(r)}\right)~({\dot{\bf r}}~\cdot {\dot{\bf
r}}~)~-~{1\over2}~\left({q\over4m}\right)^2~{\vec{\bf
\nabla}}~\left({1\over V(r)}\right)~-~q~{{\dot{\bf r}}\times {\bf r}\over
r^3} \eqno(2.10)$$
which is a complicated equation containing a velocity-squared
dependent term, typical for the motion in curved space, plus a
Coulomb term plus a Dirac-monopole term. This equation of motion
was analyzed in  literature [3-8] for the function $V(r)$ given
by eq. (1.3) which corresponds to the Kaluza-Klein monopole in
the absence of a scalar field. The analysis of eq. (2.10)  was
facilitated by the existence of some additional constant of
motion.

In what follows we shall show that  eq. (2.10) is still tractable
in spite of the complexity of the function $V(r)$  which
satisfies eq. (1.7) having one of the forms (1.8) - (1.10).

First of all we shall remark that the angular momentum
$${\bf j}~=~{\bf r}~\times~{\bf p}~+~q~{{\bf r}\over r} \eqno(2.11)$$
is conserved as in the simplifying case (1.3). Here the first
term is the orbital angular momentum and the second is the Poincar\'e
contribution which occurs when the magnetic and electric charges
are present [5]. Since these two terms are orthogonal, the
magnitude of the orbital angular momentum
$$l~=~\vert~{\bf l}~\vert~=~\vert~{\bf r}~\times~{\bf p}~\vert \eqno(2.12)$$
is also conserved. Eq. (2.11) implies that
$${\bf j}~\cdot~{{\bf r}\over r}~=~q \eqno(2.13)$$
which fixes the relative motion to lie on a cone whose vertex is
at the origin, and whose axis is ${\bf j}$.

Finally, there is a conserved vector analogous to the Runge-Lenz
vector of the Coulomb problem. Its existence is rather
surprising in view of the complexity of eq.(2.10). When the
scalar field $\Phi$ is omitted, therefore for $V(r)$ given by
eq. (1.3), this conserved vector is [5]
$${\bf K}~=~{\bf p}~\times~{\bf
j}~+~\left(~{q^2\over4m}~-~4mE~\right)~{{\bf r}\over r}~~.
\eqno(2.14) $$

Unfortunately for a function $V(r)$ as in eqs. (1.8)- (1.9)
this simple form turns out to be inadequate . In general,  for any
central potential, it is possible to construct a constant of
motion which generalizes the Runge-Lenz vector of the Coulomb
(Kepler) problem. Motivated by the study of Peres [19] we shall
construct a vector ${\bf K}$ as in eq. (2.14) with some arbitrary
functions of $r$ multiplying the vectors ${\bf p}~\times~{\bf j}$
 and  ${\bf r}$. But we prefer to avoid this approach which
implies some differential equations for these unknown functions
and to express the vector  ${\bf K}$  in terms of  $r , \theta$ and
 $\varphi$ [20].

For this purpose let us choose the $z$ axis along ${\bf j}$ so
that the motion of the particle may be conveniently described in
terms of the polar coordinates
$${\bf r}~=~r~{\bf e}~(\theta , \varphi) \eqno(2.15) $$
with
$${\bf e}~=~(\sin\theta~\cos\varphi~,~\sin\theta~\sin\varphi
{}~,~\cos\theta) \eqno(2.16) $$
and from eq. (2.13)
$${\bf j}~\cdot~{{\bf r}\over r }~=~j~\cos\theta~=~q ~~.  \eqno(2.17)$$

Having in mind that $j$ and $q$ are constants in time, eq.
(2.17) implies that $\theta$ is also constant. For the radial
variable $r$ we have from eqs. (2.7), (2.8), (2.11) and (2.12)
$$\dot{r}(r)~=~{dr\over dt}~=~\left[~2EV(r)~-~{l^2V^2(r)\over r^2}
{}~-~{q^2\over16m^2}~\right]^{1\over2} ~~.  \eqno(2.18)$$

The turning points are the roots of the equation $\dot{r}(r)~=~0$.
Assuming that at $t~=~0$ the particle starts from the turning
point $r_1$, the time dependence of the motion is given by
$$t(r)~=~\int_{r_1}^{r} {d\rho\over \dot{r}(\rho)}~~ . \eqno(2.19)$$

For the function $\varphi(r)$ it is convenient to evaluate $l^2$
and $j^2$ from eqs.(2.11) and (2.12)
$$l^2~=~{r^4\over V^2(r)}~\left(1~-~{q^2\over
j^2}~\right)~\dot{\varphi}^2 \eqno(2.20)$$
$$j^2~=~l^2~+~q^2 ~~.\eqno(2.21)$$

These equations have two solutions :

a)The angular momentum vanishes
$$l~=~0~~~,~~~j~=~q~~~,~~~\theta~=~0 ~~.\eqno(2.22)$$
In this case the energy is
$$E~=~{1\over V(r)}~\left(~\dot r^2~+~{q^2\over16m^2}~\right)~~ ,\eqno(2.23)$$
and the motion is restricted to the $j$ axis and the angle $\varphi$
is irrelevant.

b)The angular momentum is nonvanishing
$$j^2\sin^2\theta~=~l^2~~~~~,~~~\theta\not= 0$$
$$j~=~{r^2~\dot{\varphi}\over V(r)}  \eqno(2.24)$$
and consequently
$$\varphi(r)~=~j~\int_{r_1}^{r}{V(r)\over \rho^2~\dot{r}(\rho)}d\rho~~ .
\eqno(2.25)$$
Finally ,the velocity vector can be express as
$${\dot{\bf r}}~=~\dot{r}{\bf e}~+~r\dot{\varphi}{\bf e} '~
=~\dot{r}{\bf e}~+~{jV(r)\over
r}{\bf e} '\eqno(2.26)$$
where
$${\bf e} '~=~{d{\bf e}\over d\varphi}~=~(-\sin\theta~\sin\varphi~,~\sin\theta~
\cos\varphi~,~0)~~. \eqno(2.27)  $$

With these preparatives the generalized Runge-Lenz vector can be
written in a local rotating basis $({\bf e}(\varphi),~{\bf e}'(\varphi),~
{\bf j})$ as
$${\bf K}~=~X_1~\left(~{\bf e}~-~\cos\theta~{{\bf j}\over j}~\right)~+~X_2~
{\bf e}' ~~.\eqno(2.28) $$

${\bf K}$ will remain constant in the laboratory frame if it will
rotate in the opposite direction with respect to its local basis
[20]
\vfil\eject

$$ X_1~=~X_0~\cos (\varphi~-~\varphi_0~)$$
$$ X_2~=-~X_0~\sin (\varphi~-~\varphi_0~) \eqno(2.29)$$
where $X_0,~\varphi_0$ are constants which can be choosen as
one wishes.

Thus we have constructed the vector ${\bf K}$ which is constant
in time, for the motion corresponding to the lagrangian (2.6)
with an arbitrary function  $V(r)$  satisfying eq.(1.7). The
function $\varphi(r)$ occuring in its expression is given
explicitely in terms of $V(r)$ by eqs. (2.25) and (2.18).

Of course it is possible to express the generalized Runge-Lenz
vector (2.27) in a form similar to eq.(2.14) with some
functions depending of $r$ as coefficients. To put into
practice this task proves to be quite involved and by no means
illuminating.

More important it is to note that the property of being constant
in time is not sufficient for ${\bf K}$ to be an integral of the
motion. It must also be a one-valued function of the state of
the particle [20,21]. Suppose that for a set of constants  $q ,
l ,~ e~ $ in eq. (2.18), $r$ varies between the endpoints $r_1,
r_2$ which are the roots of the equation
$\dot{r}(r)~=~0$. Let us define the interval $\tau$ as time during
which $r$ varies from $r_1$ to $r_2$ and back. During time
$\tau$ the phase $\varphi$ increases  with $\Delta\varphi$. From
eq.(2.29) the sufficient condition for ${\bf K}$ to be a
one-valued function is
$$\Delta\varphi~=~2\pi k~~~~,~~~k~=~1,2,3,... \eqno(2.30)$$

The orbit of the motion satisfying the above condition form a
closed path and the period $T$ of the motion is just the
interval $\tau$. Closed paths are possible not only under the
above condition, but also in the more general case
$$\Delta\varphi~=~2\pi {k\over n} \eqno(2.31)$$
with natural numbers $k$ and $n$. In this case the period of
the motion is $T~=~n\tau$. In general, the conditions  (2.30), (2.31),
 with natural $k$ and $n$ may be met for orbits with
some particular values of $E ,~j,~l$. Therefore, contrary to
the Kepler (Coulomb) potential [21], or pure Kaluza-Klein
monopole [1] only in some particular cases the orbits are closed
paths and the vector ${\bf K}$ (2.28) may serve as an integral of
motion .
\vfil\eject
{\bf 3.CONCLUDING REMARKS}
\vskip .1in

The nonlinear sigma-model in curved space proved to be an
interesting mechanism for space-time compactification in Kaluza-Klein
theories. The compactified space becomes isomorphic to the
manifold in which the scalar fields take values and the
four-dimensional space has no cosmological term at the classical
level.

The presence of the nonlinear sigma-model coupled to gravity
does not hamper the existence of static solutions of
multi-monopole type. However, the dynamics in
this case is more involved with some notable differences . The
most important one refers to the conserved vector ${\bf K}$
analogous to the Runge-Lenz vector of the Coulomb problem. In
contrast with the pure Kaluza-Klein theory, without sources,
the generalized Runge-Lenz vector is not in general a one-valued
function of the state of the particle . For the vector ${\bf K}$
given by eq. (2.14), we have
$$\left[{\bf K}~+~{4m\over
q}\left(~E~-~{q^2\over16m^2}\right){\bf j}~\right]~\cdot~{{\bf r}\over
r}~=~j^2~-~q^2 \eqno(3.1)$$
and consequently the relative motion is in a fixed plane.
Combining this result with the fact that the relative motion is
on a cone whose axis is ${\bf j}$ one gets that  the orbits are
conic sections. When the sigma fields are present, an equation
similar to (3.1), with a constant in the right hand side,
cannot be established. The relative motion remains on a cone,
but not restricted to a plane. Moreover, in general, the orbits are not
closed paths.

The quantum dynamics of the monopoles in sourceless
five-dimensional general relativity was studied using various
methods [5,6,8,22,23]. Even in the pure Kaluza-Klein model the
quantization of the Runge-Lenz vector ${\bf K}$ is a hard task
[6,24]. We expect that the quantization of the monopole system
in the presence of a sigma-model to be more involved. The
classical trajectories are not periodic in general and new
delicate problems arise [25].

An interesting problem is to consider the motion of spinning
particles in curved space - time. We intend to extend the spaces
considered previously with additional fermionic dimensions,
parametrised by vectorial Grassmann co-ordinate $\Psi^\mu$ [26].
The relation between symmetries of the graded manifolds and
constant of motion for spinning particles is more complicated
than in the case of scalar point particles.

These generalizations will be presented elsewhere [27].
\vfil\eject

{\bf REFERENCES}

{\item ~1. R.Sorkin ,Phys.Rev.Lett.{\bf 51} (1983) 87.

          ~~~~D.J.Gross and M.J.Perry ,Nucl.Phys.{\bf B226} (1983)29 .}
{\item ~2. C.W.Misner ,J.Math.Phys.{\bf 4} (1963) 924.}
{\item ~3. N.S.Manton ,Phys.Lett.{\bf B110} (1985) 54;
 ;{\bf B154} (1985) 397 ;

{}~~~~(E){\bf B157} (1985) 475.}
{\item ~4. M.F.Atiyah and N.Hitchin ,Phys.Lett.{\bf A107} (1985) 21.}
{\item ~5. G.W.Gibbons and N.S.Manton ,Nucl.Phys.{\bf B274} (1986) 183 .}
{\item ~6. G.W.Gibbons and P.J.Ruback ,Phys.Lett.{\bf B188} (1987) 226;

{}~~~~Comm.Math.Phys.{\bf 115} (1988) ,267 .}
{\item ~7. L.Gy.Feher and P.A.Horvathy ,Phys.Lett.{\bf B182} (1987) 183
;(E){\bf B188} (1987) 512.}
{\item ~8. B.Cordani ,L.Gy.Feher and P.A.Horvathy ,Phys.Lett.{\bf B201} (1988)
 481.}
{\item ~9. C.Omero and R.Percacci ,Nucl.Phys.{\bf B165} (1980) 351.}
{\item ~10.~M.Gell-Mann and B.Zwiebach ,Phys.Lett. {\bf B141} (1984) 333 ;

{}~~~~~{\bf B147} (1985) 111 ;Nucl.Phys.{\bf B260} (1985) 569.}
{\item ~11.~S.Ianus and M.Visinescu ,Class.Quantum Gravity {\bf 3} (1986) 889.}
{\item ~12.~P.Baird and J.Eells ,Lecture Notes in Mathematics ,vol.894 (1980)

{}~~~~~(Berlin:Springer)

{}~~~~~J.Eells and L.Lemaire ,Selected Topics in Harmonic Maps ,(Conference

{}~~~~~Board of the Math.Sciences No.5) (1983) (Providence ,RI :Am.Math.Soc.)
.}
{\item ~13.~S.Ianus and M.Visinescu , Clas.Quantum Gravity {\bf
4} (1987) 1317 .}
{\item ~14.~M.Visinescu , Europh.Lett. {\bf 4} (1987) 767 .}
{\item ~15.~M.Visinescu , Europh.Lett. {\bf 10} (1989) 101 .}
{\item ~16.~M.Visinescu , Rev.Roum.Phys. {\bf 36} (11991) 633 ;

{}~~~~~~Proceedings of the XXIV International Symposium , Ahrenshoop
, (1990) ,

{}~~~~~ (Ed. G.Weight:Zeuthen , Germany ) .}
{\item ~17.~L.Bombelli , R.K.Koul , G.Kunstatter , J.Lee and
R.D.Sorkin ,

{}~~~~~ Nucl.Phys. {\bf B289} (1987) 735 ;

{}~~~~~H.M.Lee and S.C.Lee , Phys.Lett. {\bf B149} (1984) 85;

{}~~~~~F.A.Bais and P.Batenburg , Nucl.Phys {\bf B253} (1985) 162;

{}~~~~~I.G.Angus , Nucl.Phys. {\bf B264} (1986) 349 ;

{}~~~~~X.Li , F.Yu and J.Zhang , Phys.Rev. {\bf D34} (1986) 1124 ;

{}~~~~~J.Xu and X.Li , Phys.Lett. {\bf B208} (1988) 391 .}
{\item ~18.~M.Mueller and M.J.Perry , Class.Quantum Gravity {\bf
3} (1986) 65 .}
{\item ~19.~A.Peres , J.Phys.A.: Math.Gen. {\bf 12} (1979) 1711
.}
{\item ~20.~A.Holas and N.H.March , J.Phys.A.:Math.Gen. {\bf
23} (1990) 735 .}
{\item ~21.~L.D.Landau and E.M.Lifshitz ,``Mechanics`` (1976)
(Pergamon:Oxford) .}
{\item ~22.~A.Inomata and G.Junker , Phys.Lett. {\bf B234}
(1990) 41 .}
{\item ~23.~C.Grosche , J.Phys.A.:Math.Gen. {\bf 24} (1991) 1771
.}
{\item ~24.~B.Carter ,Phys.Rev. {\bf D16} (1977) 3395 .}
{\item ~25.~M.G.Gutzwiller, "The path integral in chaotic
hamiltonian systems " in

{}~~~~~~"Path integrals from meV to MeV "( 1986) (World Scientific:
     Singapore).}
{\item ~26.~F.A.Berezin and M.S.Marinov , Ann. Phys. (N.Y.) {\bf
104} (1977) 336

{}~~~~~A.Barducci, R.Casalbuoni and L.Lusanna , Nuovo Cimento {\bf
35A} (1976) 377 ;

{}~~~~~Nucl.Phys. {\bf B124} (1977) 521

{}~~~~~L.Brink, S.Deser, B.Zumino, P. Di Vecchia and P.Howe, Phys.Lett.
{\bf 64B} (1976)

{}~~~~~43

{}~~~~~R.H.Rietdijk and J.W. van Holten, Class. Quantum Gravity
{\bf 7} (1990) 247}

{\item ~27.~M.Visinescu, in preparation }

\bye